\documentstyle[preprint,aps]{revtex}
\tighten
\draft
\widetext
\input epsf
\preprint{YITP-SB-02-12, RUNHETC-2002-05}
\bigskip
\bigskip

\begin{document}
\title{On Gauge Dynamics and SUSY Breaking in Orientiworld}
\medskip
\author{Zurab Kakushadze\footnote{E-mail: 
zurab@insti.physics.sunysb.edu}}
\bigskip
\address{C.N. Yang Institute for Theoretical Physics\\ 
State University of New York, Stony Brook, NY 11794\\
and\\
Department of Physics and Astronomy, Rutgers University, Piscataway, NJ 08855}
\date{March 3, 2002}
\bigskip
\medskip
\maketitle

\begin{abstract} 
{}In the Orientiworld framework the Standard Model fields
are localized on D3-branes sitting on top of an orientifold 3-plane. The
transverse 6-dimensional space is a non-compact 
orbifold (or a more general conifold). The 4-dimensional gravity on
D3-branes is reproduced due to the 4-dimensional 
Einstein-Hilbert term induced at the quantum level. The
orientifold 3-plane plays a crucial role, 
in particular, without it
the D3-brane world-volume theories would be conformal due to the tadpole
cancellation. We study non-perturbative gauge dynamics in various 
${\cal N}=1$ supersymmetric 
orientiworld models based on the ${\bf Z}_3$ as well as ${\bf Z}_5$ and
${\bf Z}_7$ orbifold groups. Our discussions illustrate that there is a
rich variety of supersymmetry preserving dynamics in some of these models. 
On the other hand, we also find some models with dynamical supersymmetry 
breaking. 
\end{abstract}
\pacs{}

\section{Introduction}

{}Extra dimensions naturally arise in 
superstring theory (or M-theory), which is believed to be a consistent theory
of quantum gravity. 
However, in order to model the real world with 
critical string theory (or M-theory), one must
address the question of why the extra dimensions have not been observed.
One way to make extra dimensions consistent with observation is to assume
that they are compact with small enough volume. If the Standard Model gauge
and matter fields propagate in such extra dimensions (as is the case in, say,
weakly coupled heterotic string theory), then their linear sizes
should not be larger than about inverse TeV \cite{anto}. On the other hand, 
in the Brane World scenario 
the Standard Model gauge and matter fields
are assumed to be localized on branes (or an intersection thereof), 
while gravity lives in a larger dimensional bulk of space-time. Such a scenario
with compact extra dimensions can, for instance, 
be embedded in superstring theory via
Type I$^\prime$ 
compactifications. Then the extra dimensions transverse to the branes
can have sizes as large as about a tenth of a millimeter \cite{TeV}. 

{}To begin with considering compact (or, more generally, finite
volume) extra dimensions was motivated by the requirement that at the distance
scales for which gravity has been measured one should reproduce 4-dimensional
gravity. However, as was pointed out in \cite{DGP,DG}, 4-dimensional gravity
can be reproduced even in theories with infinite-volume extra dimensions.
In particular, according to \cite{DG} 4-dimensional gravity can be
reproduced on a 3-brane in infinite-volume bulk (with 6 or more space-time
dimensions) up to ultra-large distance scales. Thus, in these scenarios
gravity is almost completely localized on a brane (which is 
almost $\delta$-function-like) with ultra-light modes penetrating into the 
bulk. As was explained in \cite{DG}, this dramatic modification of gravity
in higher codimension models with infinite volume extra dimensions
is due to the Einstein-Hilbert term on the brane, which
is induced via loops of non-conformal brane matter \cite{DGP,DG}.

{}In \cite{orient} we described an explicit string theory framework for 
embedding models with infinite-volume extra dimensions. In this framework,
which we refer to as Orientiworld, the
Standard Model gauge and matter fields are localized on (a collection of)
D3-branes embedded in infinite-volume extra space. In particular, we consider
unoriented Type IIB backgrounds in the presence of some number of 
D3-branes as well as an orientifold 3-plane embedded in an orbifolded 
space-time. The D3-brane world-volume
theory in this framework is non-conformal (at least for some backgrounds with
at most ${\cal N}=1$ supersymmetry). At the quantum level we
have the Einstein-Hilbert term induced on the branes, which leads to almost
complete localization of gravity on the D3-branes. In particular, 
as was discussed in \cite{orient}, (at least in some backgrounds) up to an
ultra-large cross-over distance scale the gravitational interactions of the
Standard Model fields localized on D3-branes 
are described by 4-dimensional laws of gravity.

{}The orientiworld framework appears has a rich structure for model
building. In particular, since the extra dimensions have infinite volume, the
number of D3-branes is arbitrary. Moreover, the number of allowed orbifold
groups is infinite. Thus, {\em a priori} the orbifold group can be an
arbitrary\footnote{More precisely, there is a mild restriction on allowed
orbifold groups if we require modular invariance of the closed string sector
in the corresponding
oriented Type IIB background.} 
subgroup of $Spin(6)$, or, if we require ${\cal N}=1$ supersymmetry
to avoid bulk tachyons, of $SU(3)$.
To obtain a finite string background, we
still must impose {\em twisted} tadpole cancellation conditions. However, 
twisted tadpoles must also be canceled in compact Type IIB orientifolds. Then
the number of consistent solutions of the latter 
type is rather limited \cite{class}
as we can only have a finite number of D3-branes, and, moreover, the number of
allowed orbifold groups is also finite as they must act crystallographically
on the compact space. On the other hand, as we already mentioned, in the
orientiworld framework the number of consistent solutions is {\em infinite},
which is encouraging for phenomenologically oriented model building.

{}This richness of the orientiworld framework can be exploited to construct
various models for phenomenological applications. Thus, in \cite{orient} we 
gave a construction of an ${\cal N}=1$ supersymmetric 3-generation 
Pati-Salam model in the orientiworld framework. The purpose of this paper is
to explore non-perturbative gauge dynamics in orientiworld models. In 
particular, we discuss ${\cal N}=1$ orientiworld models based on ${\bf Z}_3$
as well as ${\bf Z}_5$ and ${\bf Z}_7$ orbifolds. The examples we study 
illustrate that various non-perturbative phenomena can be expected in 
orientiworld models including dynamical supersymmetry breaking.

{}The rest of this paper is organized as follows. In section II we review
the orientiworld framework. In section III we discuss various couplings
in unoriented Type IIB backgrounds. In section IV we discuss the ${\bf Z}_3$
models. In section V we give the ${\bf Z}_5$ and ${\bf Z}_7$ examples. We
give some concluding remarks in section VI.

\section{Orientiworld Framework}

{}In this section we review the orientiworld framework. First we describe the
underlying oriented Type IIB orbifold backgrounds. We then consider their
orientifolds. Parts of our discussion here will closely 
\cite{BKV,orient1,orient}.

\subsection{Oriented Backgrounds}

{}Consider Type IIB string theory with $N$ parallel coincident D3-branes where
the space transverse to the
D-branes is ${\cal M}={\bf R}^6/\Gamma$. The orbifold group
$\Gamma= \left\{ g_a \mid a=1,\dots,|\Gamma| \right\}$ ($g_1=1$)
must be a finite discrete subgroup of $Spin(6)$ (it can be a subgroup of
$Spin(6)$ and not $SO(6)$ as we are dealing with a theory containing fermions).
If $\Gamma\subset SU(3)$ ($SU(2)$), we have
${\cal N}=1$ (${\cal N}=2$) unbroken supersymmetry,
and ${\cal N}=0$, otherwise.

{}Let us confine our attention to the cases where type IIB on ${\cal M}$ 
is a modular invariant 
theory\footnote{This is always the case if $\Gamma\subset SU(3)$. For 
non-supersymmetric cases this is also true provided that
$\not\exists{\bf Z}_2\subset\Gamma$. If $\exists{\bf Z}_2\subset\Gamma$,
then modular invariance requires that the set of points in ${\bf R}^6$
fixed under the ${\bf Z}_2$ twist has real dimension 2.}. The action of the
orbifold on
the coordinates $X_i$ ($i=1,\dots,6$) on ${\cal M}$ can be described
in terms of $SO(6)$ matrices:
$g_a:X_i\rightarrow \sum_j (g_a)_{ij} X_j$. (The action of $g_a$ on the
world-sheet superpartners of $X_i$ is the same.)
We also need to specify
the action of the orbifold group on the Chan-Paton charges carried by the
D3-branes. It is described by $N\times N$ matrices $\gamma_a$ that
form a representation of $\Gamma$. Note that $\gamma_1$ is an identity
matrix and ${\mbox {Tr}}(\gamma_1)=N$.

{}The D-brane sector of the theory is described by an oriented open
string theory. In particular, the world-sheet expansion corresponds
to summing over oriented Riemann surfaces with arbitrary genus $g$ and
arbitrary number of boundaries $b$, where the boundaries of the world-sheet 
correspond to the D3-branes. 
In \cite{BKV} it was shown that the one-loop massless 
(and, in non-supersymmetric 
cases, tachyonic) tadpole cancellation conditions require that
\begin{equation}\label{tadpole}
 {\mbox {Tr}}(\gamma_a)=0~~~\forall a\not=1~.
\end{equation}
In \cite{BKV} 
it was also shown that this condition implies that the Chan-Paton 
matrices $\gamma_a$
form an $n$-fold copy of the {\em regular} representation of $\Gamma$. 
The regular representation decomposes into a direct sum of all irreducible
representations ${\bf r}_i$ of $\Gamma$ with degeneracy factors
$n_i=|{\bf r}_i|$. The gauge group is ($N_i\equiv nn_i$)
\begin{equation}
 G=\otimes_i U(N_i)~. 
\end{equation}
The matter consists of Weyl fermions and scalars transforming in
bifundamentals $({\bf N}_i,{\overline {\bf N}}_j)$ 
(see \cite{LNV} for details). The overall center-of-mass $U(1)$, which is 
inherited from the parent ${\cal N}=4$ supersymmetric 
$U(N)$ gauge theory and is
always present in such models, is free - 
matter fields are not charged under this $U(1)$. We do, however, have matter
charged under the rest of the $U(1)$'s, which we will refer to as non-trivial
$U(1)$'s. If $\Gamma\subset SO(3)$, then all the non-trivial $U(1)$'s 
are anomaly free, in fact, in these cases gauge theories are necessarily 
non-chiral. As was discussed in \cite{orient},
these $U(1)$'s acquire masses at the one-loop level via couplings to
the corresponding twisted R-R two-forms for which there are induced 
kinetic terms on D3-branes.
If $\Gamma \subset SU(3)$ but $\Gamma\not\subset SO(3)$, then we have chiral
gauge theories and 
some of the non-trivial $U(1)$ factors are actually anomalous 
(in particular, we have
$U(1)_k SU(N_l)^2$ mixed anomalies), and are broken (that is, acquire masses)
at the tree-level
via a generalized Green-Schwarz mechanism \cite{IRU,Poppitz,orient}. 
If in these
cases we also have anomaly-free non-trivial $U(1)$'s, 
then the latter acquire masses at
the one-loop level just as in the $\Gamma\subset SO(3)$ cases \cite{orient}.
So the non-trivial $U(1)$ factors decouple in the infra-red. 
As to the non-Abelian
parts of the gauge theories, it was shown in \cite{BKV} that they are
conformal in the large $N$ 
limit\footnote{In this limit we take the string coupling $g_s\rightarrow 
0$ together with $N\rightarrow \infty$ while keeping $Ng_s$ fixed.}, including
in the non-supersymmetric cases. The key reason
for this conformal property is the tadpole cancellation condition 
(\ref{tadpole}), which, as was explained in \cite{BKV}, implies that all
planar diagrams\footnote{In the large $N$ limit non-planar diagrams are 
suppressed by powers of $1/N$.} 
with external lines corresponding to non-Abelian gauge as well
as matter fields reduce to those of the parent ${\cal N}=4$ theory, which is
conformal. At finite $N$ conformality of the non-Abelian parts of the
${\cal N}=2$
gauge theories becomes evident from vanishing of the one-loop $\beta$-function 
(as ${\cal N}=2$ gauge
theories perturbatively are not renormalized beyond one loop), 
and can also be argued for
${\cal N}=1$ cases \cite{BKV}.
In non-supersymmetric cases, however, we always have twisted closed string 
sectors tachyons, which 
prevent one from considering finite $N$ cases\footnote{In the large $N$
limit the closed twisted sector tachyons are harmless as the string coupling 
$g_s$ goes to zero. Also, in this limit all non-trivial $U(1)$'s
decouple in the infra-red.}.

\subsection{Unoriented Backgrounds}

{}Let us now consider a generalization of the above setup by including
orientifold planes. In the following we will mostly be interested in finite $N$
theories, so let us focus on theories with at least ${\cal N}=1$ unbroken
supersymmetry. Thus,
consider Type IIB string theory on ${\cal M}={\bf C}^3/\Gamma$ where
$\Gamma\subset SU(3)$. Consider the $\Omega (-1)^{F_L}J$ orientifold of this 
theory, where $\Omega$ is the world-sheet parity reversal, $F_L$ is the 
fermion number operator, and $J$ 
is a ${\bf Z}_2$ element ($J^2=1$) acting on the complex coordinates $z_i$
($i=1,2,3$) on ${\bf C}^3$ such that the set of points in 
${\bf C}^3$ fixed under 
the action of $J$ has real dimension $\Delta=0$ or $4$. 

{}If $\Delta=0$ then we have an orientifold 3-plane. If $\Gamma$ has
a ${\bf Z}_2$ subgroup, then we also have an orientifold 7-plane.
If $\Delta=4$ then we have an orientifold 7-plane. We may also have
an orientifold 3-plane depending on whether $\Gamma$ has an appropriate
${\bf Z}_2$ subgroup. Regardless of whether we have an orientifold 3-plane,
we can {\em a priori} introduce an arbitrary number of 
D3-branes\footnote{In general, codimension-3 and higher objects (that is
D-branes and orientifold planes) do not introduce untwisted tadpoles.}. 
On the other hand, if we have an orientifold 7-plane we must 
introduce 8 of the corresponding D7-branes to cancel the corresponding
R-R charge 
appropriately. (The number 8 of D7-branes is required by the corresponding 
untwisted tadpole cancellation conditions.) 

{}We need to specify the action of $\Gamma$ on the Chan-Paton factors
corresponding to the D3- and D7-branes (if the latter are present, which is
the case if we have an orientifold 7-plane). Just as in the previous 
subsection, 
these are given by Chan-Paton matrices which we collectively refer to
as $\gamma^\mu_a$, where the superscript $\mu$ refers to the corresponding
D3- or D7-branes. Note that ${\mbox{Tr}}(\gamma^\mu_1)=n^\mu$ where 
$n^\mu$ is the number of D-branes labelled by $\mu$. 

{}Now the world-sheet expansion contains oriented as well as unoriented
Riemann surfaces. The unoriented Riemann surfaces contain handles and 
boundaries as well as cross-caps. The latter are 
the (coherent Type IIB) states 
that describe the familiar orientifold planes. The presence of the cross-caps
modifies the twisted one-loop tadpole cancellation conditions, which can now
be written as:
\begin{equation}
 B_a+\sum_\mu C^\mu_a {\mbox{Tr}}(\gamma^\mu_a)=0~,~~~a\not=1~.
\end{equation}
These should be contrasted with (\ref{tadpole}) in the oriented case.
In particular, in certain cases some $B_a$, which correspond to
contributions due to cross-caps, need not vanish. This makes possible 
(albeit does not guarantee - see below) non-conformal gauge theories on 
D3-branes in the orientiworld context \cite{orient1,orient2,orient}.

{}Thus, let us see what kind of orientiworld models we can have. 
For definiteness
let us focus on the cases where we do have an orientifold 3-plane (that is,
$\Delta=0$). If there are
no orientifold 7-planes (that is, if $\Gamma$ does not contain a ${\bf Z}_2$
element), then the orientifold projection $\Omega$ can be either of the $SO$
or the $Sp$ type: the corresponding orientifold 3-plane is referred to as
O3$^-$ or O3$^+$, respectively. That is, before orbifolding, if we place
$2N$ D3-branes on top of the O3$^-$-plane (O3$^+$-plane), we have the ${\cal
N}=4$ super-Yang-Mills theory with the $SO(2N)$ ($Sp(2N)$) gauge 
group\footnote{Note that we can also place $2N+1$ D3-branes on top of
the O3$^-$-plane to obtain the $SO(2N+1)$ gauge group.}.
(We are using the convention where $Sp(2N)$ has rank $N$.) After the orbifold
projections the 33 (that is, the D3-brane) gauge group is a subgroup
of $SO(2N)$ ($Sp(2N)$), which can contain $U(N_k)$ factors as well as 
$SO$ ($Sp$) subgroups. The 33 matter can contain bifundamentals in any of these
subgroups as well as rank-2 antisymmetric (symmetric) representations in
the unitary subgroups. Next, if we have an O7-plane, the orientifold 
projection $\Omega$ 
must always be of the $SO$ type on the D7-branes - this is required
by the tadpole cancellation condition. This, in particular, implies that the
33 and 77 matter cannot contain rank-2 symmetric representations. Note that
we also have 37 matter in bifundamentals of the 33 and 77 gauge groups. 
Finally, note that there is no overall center-of-mass $U(1)$ in these models 
(the parent theory is an $SO$ or $Sp$ gauge theory), so all $U(1)$'s (if
present) are non-trivial in the sense of the previous subsection.

{}If $\Gamma\subset SO(3)$, then the corresponding gauge theories are
necessarily non-chiral. If we have $U(1)$ factors, they acquire masses
at the one-loop level via the mechanism discussed in the previous subsection
\cite{orient}. As to the non-Abelian parts of the corresponding gauge theories,
as was discussed in \cite{orient}, they are always conformal despite the fact
that some twisted $B_a$ can be non-zero in such models (see \cite{orient} for
a detailed explanation of why this is so)\footnote{The $\Gamma\subset SU(2)$
orientifolds were originally discussed in \cite{orient1,FS,PU}.}. 
The situation is very different
in the cases where $\Gamma\subset SU(3)$ but $\Gamma\not\subset SO(3)$. All 
such theories are non-conformal \cite{orient}. In fact, generically they are
chiral with a few (essentially trivial) exceptions. Thus, in some cases
the twisted tadpole cancellation conditions allow a choice such that
all twisted Chan-Paton matrices are trivial (that is, they are identity
matrices). In such a case the gauge theory is a pure
$SO$ or $Sp$ ${\cal N}=1$ super-Yang-Mills theory (that is, we have no 
chiral matter supermultiplets). In all other cases we have chiral matter.
More precisely, there is one possible exception where the gauge group is
$SU(4)\otimes U(1)_A$ with matter transforming in ${\bf 6}(+2)$ (the $U(1)_A$
charge is given in parenthesis). Such a theory is conformal as the matter
is non-trivially charged under the anomalous $U(1)_A$, but the latter is
broken at the tree level via a generalized Green-Schwarz mechanism, and
the resulting non-Abelian gauge theory turns out to be non-conformal 
(as ${\bf 6}$ of $SU(4)$ is a real representation). In general, if we have
chiral matter, we have at least one anomalous $U(1)$. Such anomalous $U(1)$'s
acquire masses at the tree level \cite{IRU,Poppitz,orient}. 
If in these cases we also have anomaly-free
$U(1)$'s, they acquire masses at the one-loop level as in the $\Gamma\subset
SO(3)$ cases \cite{orient}. 

\section{Various Couplings In Orientiworld}

{}For our subsequent discussions it will be useful to understand some
couplings in the orientifold backgrounds. For simplicity we will focus on the
cases with O3-planes but without O7-planes. In fact, we will specialize on
orbifold groups $\Gamma={\bf Z}_p$ such that $\Gamma\subset SU(3)$ but
$\Gamma\not\subset SO(3)$, where $p$ is a prime. The action of the generator
$\theta$ or ${\bf Z}_p$ on the complex coordinates $z_\alpha$, $\alpha=1,2,3$,
on ${\bf C}^3/\Gamma$ is given by
\begin{equation}
 \theta z_\alpha=\omega^{\ell_\alpha} z_\alpha~,
\end{equation}
where $\omega\equiv \exp(2\pi i/p)$, and $\sum_{\alpha=1}^3 \ell_\alpha=p$.

\subsection{Oriented Backgrounds}

{}Before we turn to the unoriented backgrounds, let us recall some facts about 
the oriented theories. Thus, consider Type IIB on ${\bf R}^{1,3}\times
({\bf C}^3/{\bf Z}_p)$. The closed string sector has $p-1$ twisted sectors
$\theta^k$, $k=1,\dots,p-1$.
In each twisted sector we have a complex NS-NS scalar $\phi_k$ and a complex
R-R two-form $C_k$, which satisfy the reality condition \cite{orbifold} 
\begin{equation}
 \phi_{p-k}=\phi_k^*~,
\end{equation}
and similarly for $C_k$.

{}Next, consider $N$ D3-branes placed at the orbifold fixed point in 
$({\bf C}^3/{\bf Z}_p)$. Let ${\cal F}$ be the ($N\times N$ matrix valued)
D3-brane gauge field strength, which satisfies the orbifold projection
$\gamma_k {\cal F}\gamma_k^{-1}={\cal F}$ (recall that we have $N=np$ and
$\gamma_k={\rm diag}(I_n, \omega I_n, \omega^2 I_n,\dots,\omega^{p-1}I_n)$,
where $I_n$ is the $n\times n$ identity matrix). 
We have a Chern-Simons coupling of the
following form \cite{Doug,DM,DGM}:
\begin{equation}\label{CS}
 S_{\rm \small{CS}}={1\over 2\pi\alpha^\prime}\sum_{k=1}^{p-1} 
 \int_{\rm D3} C_k 
 \wedge {\rm Tr}\left(\gamma_k ~e^{2\pi\alpha^\prime{\cal F}}\right)~.
\end{equation}  
In particular, the term linear in ${\cal F}$
\begin{equation}\label{mixing}
 \sum_{k=1}^{p-1} \int_{\rm D3} C_k 
 \wedge {\rm Tr}\left(\gamma_k ~{\cal F}\right)
\end{equation}
describes the mixing between the $p-1$ twisted two-forms $C_k$ and $p-1$
anomalous $U(1)$'s (note that we have $p$ $U(1)$'s, but one of them is an
overall center-of-mass $U(1)$ which does not couple to the 
twisted two-forms $C_k$). Since the fields $C_k$ have non-vanishing kinetic 
terms supported at the orbifold fixed point (that is, they propagate in ${\bf
R}^{1,3}$ that coincides with the D3-brane world-volumes), the anomalous
$U(1)$'s are actually massive already at the tree level.

{}Now consider the supersymmetric completion of the couplings 
(\ref{mixing}), which gives the corresponding Fayet-Iliopoulos (FI) couplings:
\begin{equation}
 S_{\rm {\small FI}}=\sum_{k=1}^{p-1} \int_{\rm D3} \phi_k 
 {\rm Tr}\left(\gamma_k ~{\cal D}\right)~,
\end{equation}
where ${\cal D}$ is the (matrix valued) auxiliary field corresponding to 
${\cal F}$. The FI term for a given non-trivial $U(1)_j$, $j=1,\dots,p-1$, 
therefore reads:
\begin{equation}
 \xi_{{\rm{\small FI}},j}=\sum_{k=1}^{p-1} \phi_k {\rm Tr}\left(\gamma_k ~
 \lambda_j\right)~,
\end{equation}
where $\lambda_j$ is the Chan-Paton matrix corresponding to this $U(1)_j$.
Note that at the orbifold point the D-terms give masses to the twisted
NS-NS scalars $\phi_k$, which are now part of the massive $U(1)$ gauge
supermultiplets. 

{}Next, consider the terms in (\ref{CS}) quadratic in ${\cal F}$: 
\begin{equation}
 \pi \alpha^\prime\sum_{k=1}^{p-1} \int_{\rm D3} C_k 
 \wedge {\rm Tr}\left(\gamma_k ~{\cal F}^2\right)~.
\end{equation}
The supersymmetric completion of this coupling gives a coupling proportional
to
\begin{equation}\label{gauge}
 \sum_{k=1}^{p-1} \int_{\rm D3} \phi_k {\rm Tr}
 \left(\gamma_k ~{\cal F}^2\right)~.
\end{equation}
That is, the twisted NS-NS scalars contribute to the gauge couplings,
while the twisted R-R scalars (dual to the twisted two-forms) contribute
to the corresponding $\theta$-angles (that is, the gauge kinetic function
is given by $f=S+f_1$, where $S$ is the untwisted sector 
dilaton supermultiplet,
while $f_1$ is the contribution which depends on the twisted closed
string sector moduli) \cite{LNV,IRU}.

\subsection{Unoriented Backgrounds}

{}Once we add an O3-plane, the above discussion is modified as follows.
First, note that the twisted Chan-Paton matrices now have the form:
$\gamma_k={\rm diag}(I_{n_0},\omega I_{n_1},\dots, \omega^{p-1} I_{n_{p-1}})$,
where $\sum_{k=0}^{p-1} n_k=N$, and the integers $n_k$ satisfy the reality 
condition $n_{p-k}=n_k$, $k=1,\dots,p-1$, and otherwise are no longer identical
but are determined by the twisted tadpole cancellation conditions
\cite{KaSh,orient1,orient2,IRU} (here $\eta=-1$ for the O3-$^{-}$ plane,
while $\eta=1$ for the O3$^+$-plane):
\begin{equation}
 {\rm Tr}\left(\gamma_{2k}\right)=-4\eta\prod_{\alpha=1}^3 
 \left(1+\omega^{k\ell_\alpha}\right)^{-1}~.
\end{equation}
In particular, some $n_k$ can actually vanish. The
gauge group is now $SO(n_0)$ or $Sp(n_0)$ (depending on whether we choose an
O3$^-$- or O3$^+$-plane) times $\bigotimes_{k=1}^{(p-1)/2} U(n_k)$ (if any of 
the $n_k$ vanishes, we simply delete the corresponding subgroup).
Second, the orientifold projection
removes the real parts of the complex fields $\phi_k$ and $C_k$, while the
imaginary parts ${\rm Im}(\phi_k)$ and ${\rm Im}(C_k)$, $k=1,\dots,(p-1)/2$,
are combined (after
dualizing the two-forms $C_k$ to scalars ${\widetilde\phi}_k$)
into $(p-1)/2$ twisted chiral supermultiplets
\cite{DGM,CVET}. If we 
actually have an anomalous $U(1)_{n_j}$ factor coming from the 
$U(n_j)$ subgroup
(that is, if the corresponding $n_j\not=0$), then this $U(1)_{n_j}$ 
becomes massive
at the tree level via the Chern-Simons coupling of the corresponding field
strength ${\cal F}_{n_j}$ with the fields ${\rm Im}(C_k)$. 
This can be seen from the
part of (\ref{mixing}) surviving the
orientifold projection (it is not difficult to see that the Chan-Paton matrix
$\lambda_{n_j}$ for this $U(1)_{n_j}$ factor is a matrix 
block-diagonal w.r.t. partitions
of $N$ into $n_0,\dots,n_{p-1}$ integers with only non-vanishing entries
being $n_j$ 1's and $n_{p-j}(=n_j)$ $-1$'s): 
\begin{equation}
 -2 \sum_{k=1}^{(p-1)/2}\sum_{j=1}^{(p-1)/2} {\rm Im}\left({\rm Tr}
 \left[\gamma_k\lambda_{n_j}\right]\right) \int_{\rm D3} {\rm Im}(C_k) \wedge 
 {\cal F}_{n_j}~.
\end{equation}
Similarly, the Fayet-Iliopoulos terms are given by:
\begin{equation}\label{FI}
 \xi_{{\rm{\small FI}},n_j}=-2\sum_{k=1}^{(p-1)/2} {\rm Im}(\phi_k)~
 {\rm Im}\left({\rm Tr}
 \left[\gamma_k\lambda_{n_j}\right]\right)~.
\end{equation}
These couplings imply that, since we have $m_A=\sum_{j=1}^{(p-1)/2}\left[1-
\delta_{n_j,0}\right]$ anomalous $U(1)$'s, precisely $m_A$ linear combinations
of the $(p-1)/2$ chiral superfields (whose lowest components are given by 
complex scalars $\Phi_k={\rm Im}(\phi_k)+i~{\rm Im}({\widetilde \phi}_k)$)
become part of the massive $U(1)$ superfields at the orbifold point.  

{}Finally, let us see what happens to the correction $f_1$ to the gauge kinetic
function due to the twisted moduli $\Phi_k$. In the unoriented backgrounds
these corrections actually {\em vanish} \cite{CVET}. Thus, consider the part of
(\ref{gauge}) surviving the orientifold projection. For the $SO(n_0)/Sp(n_0)$
part of the gauge group (if present) the corresponding part of the 
trace ${\rm Tr}\left(\gamma_k {\cal F}^2\right)$ is always real, so due to
the reality condition $\phi_{p-k}=\phi_k^*$ only the real parts of $\phi_k$
can contribute. But it is precisely the real parts of $\phi_k$ that are 
removed 
by the orientifold projection. As to the $U(n_j)$ subgroups, note that up to
the appropriate normalization factors the corresponding corrections to $f_1$
are the same for the non-Abelian parts $SU(n_j)$ as those for the Abelian 
parts $U(1)_{n_j}$. The latter are given by
\begin{equation}
 \sum_{k=1}^{p-1}\sum_{j=1}^{(p-1)/2}{\rm Tr}
 \left(\gamma_k\lambda_{n_j}^2\right)
 \int_{\rm D3}\phi_k{\cal F}_{n_j}^2~.
\end{equation}
Note that the traces ${\rm Tr}\left(\gamma_k\lambda_{n_j}^2\right)$
are all real, so only the real parts of $\phi_k$ can contribute. This implies 
that in the orientifold backgrounds the gauge couplings (and the corresponding
$\theta$-angles) do not receive twisted moduli dependent corrections 
at the tree level\cite{CVET}.

\section{The ${\bf Z}_3$ Models}

{}The simplest choice of the orbifold group $\Gamma$ in the present context
is $\Gamma={\bf Z}_3=\{1,\theta,\theta^2\}$, where the generator $\theta$ of
${\bf Z}_3$ acts on the complex coordinates $z_\alpha$, $\alpha=1,2,3$, 
on ${\bf C}^3$ as follows:
\begin{equation}
 \theta z_\alpha=\omega z_\alpha~,
\end{equation}
where $\omega\equiv\exp(2\pi i/3)$. At the origin of ${\bf C}^3/\Gamma$ we can
place an O3$^-$- or O3$^+$-plane. Let $\eta=-1$ in the former case, while
$\eta=+1$ in the latter case. The twisted tadpole cancellation requires that
\cite{orient1}
\begin{equation}
 {\rm Tr}(\gamma_\theta)=4\eta~.
\end{equation} 
We can then place $(3N+4\eta)$ D3-branes on top of the O3-plane, and choose
\begin{equation}
 \gamma_\theta={\rm diag}(\omega I_N,\omega^{-1} I_N, I_{N+4\eta})~. 
\end{equation}
The twisted closed string sector gives rise to a single massless chiral 
supermultiplet corresponding to the orbifold blow-up mode.
The massless open string spectrum gives rise to the gauge theory on the 
D3-branes. For $\eta=1$ and $N=0$ we have pure $Sp(4)$ super-Yang-Mills 
theory\footnote{Recall that in 
our conventions $Sp(N)$ with $N\in 2{\bf N}$ has rank 
$N/2$.}. For $\eta=1$ and $N\in 2{\bf N}$ we have $SU(N)\otimes Sp(N+4)\otimes
U(1)_A$ gauge theory with matter in the chiral supermultiplets
$\Phi_\alpha=3\times ({\bf S},{\bf 1})(+2)$ and 
$Q_\alpha=3\times ({\overline {\bf N}},
{\bf N+4})(-1)$, where the anomalous $U(1)_A$ charges are given in 
parentheses, and ${\bf S}$ stands for the two-index $N(N+1)/2$ dimensional 
symmetric representation of $SU(N)$. For $\eta=-1$ and $N=4$ we have 
$SU(4)\otimes U(1)_A$ gauge theory with matter in the chiral supermultiplets 
$\Phi_\alpha=3\times {\bf 6}(+2)$. For 
$\eta=-1$ and $N=5$ we have $SU(5)\otimes
U(1)_A$ gauge theory with matter in the chiral supermultiplets 
$\Phi_\alpha=3\times {\bf 10}(+2)$ and ${\bf Q}_\alpha=3\times 
{\overline {\bf 5}}(-1)$. For
$\eta=-1$ and $N\geq 6$ we have $SU(N)\otimes SO(N-4)\otimes
U(1)_A$ gauge theory with matter in the chiral supermultiplets 
$\Phi_\alpha=3\times ({\bf A},{\bf 1})(+2)$ and ${\bf Q}_\alpha=3\times
({\overline {\bf N}},
{\bf N-4})(-1)$, where ${\bf A}$ stands for the two-index $N(N-1)/2$ 
dimensional antisymmetric 
representation of $SU(N)$. (Note that in the $\eta=-1$ 
and $N=6$ case the $SO$ part of the gauge group is actually Abelian.)
In the cases where we have the $Q_\alpha$ matter, we have the following
tree-level superpotential:
\begin{equation}\label{tree}
 {\cal W}_{\rm{\small tree}}=y \epsilon_{\alpha\beta\gamma} 
 \Phi_\alpha Q_\beta Q_\gamma+\dots~,
\end{equation}
where $y$ is the corresponding Yukawa coupling, and 
the ellipses stand for non-renormalizable couplings. Note that at the 
renormalizable level we have an $SO(3)$ global 
symmetry\footnote{Non-renormalizable couplings suppressed by powers of $M_s$ 
break this global $SO(3)$ symmetry to its discrete subgroup subsumed in the
discrete ${\bf Z}_3$ gauge symmetry.}, and the subscript
$\alpha$ in $\Phi_\alpha$ and $Q_\alpha$ corresponds to the triplet of $SO(3)$.

\subsection{The $Sp$ Theories}

{}Some examples (more precisely, their compact versions) of the above 
theories with $\eta=-1$ (that is, with the $SO$ orientifold projection) 
were discussed in \cite{Sagnotti,KaSh,KST,LPT,CVET}. Here we will focus on
the theories with $\eta=1$ (that is, with the $Sp$ orientifold projection).
Let us note that the $\eta=1$ models are actually simpler to discuss
then their $\eta=-1$ counterparts. This is because in the $\eta=-1$ cases
the $SU$ part of the gauge theory is asymptotically free 
(while the $SO$ part is not asymptotically free), 
so in the infra-red we have to deal with a strongly 
coupled chiral gauge theory\footnote{One exception is the $\eta=-1$
and $N=4$ model, where the non-Abelian gauge group is $SU(4)$, the matter
consists of 3 chiral supermultiplets in ${\bf 6}$ of $SU(4)$, and there is 
no tree-level superpotential. This theory can be thought of as $SO(6)$ with
3 flavors, which is well understood \cite{IS}.}. In contrast, in the $\eta=1$
case it is the $SU$ part that is not asymptotically free, while the $Sp$ part
is, so in the infra-red the non-perturbative dynamics reduces to that of 
a strongly coupled $Sp(N_c=N+4)$ gauge theory with $N_f=3N/2$ flavors
(by one flavor of $Sp(N_c)$ we mean a pair of chiral supermultiplets in
$N_c$ of $Sp(N_c)$). And these theories are well understood \cite{IP}.

{}First, consider the model with $\eta=1$ and $N=0$. The gauge theory on the
D3-branes is pure $Sp(4)$ super-Yang-Mills theory. Note that the twisted closed
string sector gives rise to a single chiral supermultiplet, which plays no
role in the gauge dynamics (this follows from our discussion in the 
previous section). Non-perturbatively we have gaugino condensate in $Sp(4)$
with confinement and chiral symmetry breaking. This model is interesting
as it provides a simple setup where one gets pure ${\cal N}=1$ superglue
on D3-branes. In particular, it would be interesting to use this setup
to identify BPS domain walls in the $Sp(4)$ super-Yang-Mills theory, but
this is outside of the scope of this paper.

{}Let us now discuss the $\eta=1$ and $N\in 2{\bf N}$ models. For 
presentation purposes we will discuss the $N=2$ model after we discuss
all the other cases.

\begin{center}
 {\em The $SU(4)\otimes Sp(8)\otimes U(1)_A$ Model}
\end{center} 

{}This is the $\eta=1$ and $N=4$ model. The gauge group is 
$SU(4)\otimes Sp(8)\otimes U(1)_A$, the chiral matter is given by 
$\Phi_\alpha=3\times ({\bf 10},{\bf 1})(+2)$ and 
$Q_\alpha=3\times ({\overline {\bf 4}},
{\bf 8})(-1)$, and the tree-level superpotential is given by (\ref{tree}).
As we have already mentioned, the $SU(4)$ gauge coupling is weak in the 
infra-red, while the $Sp(8)$ gauge coupling becomes strong. The low energy
degrees of freedom are given by mesons (note that there are no baryons in
$Sp$ theories \cite{IP})
\begin{eqnarray}
 &&{\cal M}_{[\alpha\beta]}=3\times({\overline {\bf 10}},{\bf 1})(-2)~,\\
 &&{\cal M}_{\{\alpha\beta\}}=6\times({\overline {\bf 6}},{\bf 1})(-2)~.
\end{eqnarray}
(Note that for $SU(4)$ we actually have ${\overline {\bf 6}}={\bf 6}$.)
There is no non-perturbative superpotential in this theory, and at the origin
of the meson moduli space we have confinement without chiral symmetry breaking
\cite{IP}. Note that due to the tree-level superpotential (\ref{tree}) 
the mesons ${\cal M}_{[\alpha\beta]}$ pair up with the fields $\Phi_\alpha$
and acquire masses:
\begin{equation}
 {\cal W}_{\rm{\small tree}}=y\epsilon_{\alpha\beta\gamma}\Phi_\alpha
 {\cal M}_{[\beta\gamma]}~.
\end{equation}
So at low energies we have the $SU(4)$ gauge theory with matter chiral 
supermultiplets in $P_I=6\times {\overline {\bf 6}}(-2)$, $I=1,\dots,6$. 
Actually, so far we have
been ignoring the anomalous $U(1)_A$. The corresponding D-term is given by
\begin{equation}
 D=-2P^2+\xi_{\rm{\small FI}}~.
\end{equation}
{}From (\ref{FI}) it follows that $\xi_{\rm{\small FI}}$ is negative for
positive values of ${\rm Im}(\phi_1)$ (which corresponds to the size
of the blow-up). This implies that all the fields $P_I$ as well as  
${\rm Im}(\phi_1)$ have vanishing expectation values. That is, the blow-up
mode is frozen and the orbifold {\em cannot} be blown up in this model.
As we discussed in the previous section, at the orbifold point the twisted
chiral superfield becomes part of the massive $U(1)_A$ gauge supermultiplet.
The low energy theory is therefore $SO(6)$ gauge theory with 6 vectors. Note 
that this theory is not infra-red free. However, it has a dual magnetic
description \cite{IS} in terms of $SO(4)\sim SU(2)_L\otimes SU(2)_R$ gauge 
theory with 6 flavors of quarks $q^I$ ($I=1,\dots,6$) in 
the $4\sim (2,2)$ dimensional representation along with $6(6+1)/2=21$ gauge
singlets $M_{IJ}$ and the superpotential
\begin{equation}
 {\cal W}_{\rm{\small magnetic}}=M_{IJ}q^I q^J~.
\end{equation} 
Note that the magnetic theory is actually free in the infra-red.

\begin{center}
 {\em The $SU(N)\otimes Sp(N+4)\otimes U(1)_A$ Models with $N>4$}
\end{center}

{}Before we discuss the $\eta=1$ and $N=6$ model, we would like to discuss
the $\eta=1$ and $N>6$ models. In this case the $Sp(N+4)$ theory has a dual
magnetic description (albeit the magnetic theory is also strongly coupled)
\cite{IP}. In this dual picture the gauge group is $SU(N)\otimes Sp(2N-8)
\otimes U(1)_A$, the matter is given by
\begin{eqnarray}
 &&\Phi_\alpha=3\times ({\bf S},{\bf 1})(+2)~,\\
 &&{\cal M}_{[\alpha\beta]}=3\times ({\overline {\bf S}},{\bf 1})(-2)~,\\
 &&{\cal M}_{\{\alpha\beta\}}=6\times ({\overline{\bf A}},{\bf 1})(-2)~,\\
 &&q_\alpha=3\times ({\bf N},{\bf 2N-8})(+1)~, 
\end{eqnarray} 
and the superpotential is given by
\begin{equation}
 {\cal W}_{\rm{\small magnetic}}=y\epsilon_{\alpha\beta\gamma}\Phi_\alpha
 {\cal M}_{[\beta\gamma]}+{\cal M}_{\{\alpha\beta\}}q_\alpha q_\beta~.
\end{equation}
The mesons ${\cal M}_{[\alpha\beta]}$ pair up with the fields $\Phi_\alpha$ and
acquire masses, so at low energies we have the mesons 
${\cal M}_{\{\alpha\beta\}}$ and the quarks $q_\alpha$. As we have already
mentioned, the $Sp(2N-8)$ gauge coupling is strong in the infra-red. On the
other hand, the $SU(N)$ gauge coupling remains weak (even after integrating out
the fields ${\cal M}_{[\alpha\beta]}$ and $\Phi_\alpha$). Moreover, the
orbifold blow-up mode is frozen in this model. This can be seen as follows.
The $U(1)_A$ D-term reads:
\begin{equation}
 D=-2M_{\{\alpha\beta\}}^2 +q^2+\xi_{\rm{\small FI}}~,
\end{equation}
where $\xi_{\rm{\small FI}}<0$ for positive ${\rm Im}(\phi_1)$. We must
also ensure D-flatness for the $SU$ and $Sp$ subgroups. The D-flat
directions are in one-to-one correspondence with chiral gauge invariant 
operators. As far as the $Sp$ part is concerned, such operators can only
contain the following combinations:
\begin{eqnarray}
 &&\Sigma_{[\alpha\beta]}=
 q_{[\alpha}q_{\beta]}=3\times ({\bf S},{\bf 1})(+2)~,\\
 &&\Sigma_{\{\alpha\beta\}}=q_{\{\alpha}q_{\beta\}}=
 6\times ({\bf A},{\bf 1})(+2)~.
\end{eqnarray}
Note, however, that $\Sigma_{[\alpha\beta]}$ cannot enter as we cannot 
construct $SU(N)$ gauge invariant operators from $\Sigma_{[\alpha\beta]}$,
$\Sigma_{\{\alpha\beta\}}$ and $M_{\{\alpha\beta\}}$. On the other hand,
the F-flatness conditions imply that $\Sigma_{\{\alpha\beta\}}=0$. It then 
follows that the FI term must vanish along with the vacuum expectation values
of ${\cal M}_{\{\alpha\beta\}}$ and $q_\alpha$.

{}Let us now discuss the $N=6$ case. The above discussion is essentially
unmodified in this case except that in the dual magnetic theory the $Sp(4)$
subgroup is actually weakly coupled. As to the $SU(6)$ gauge theory, its
one-loop $\beta$-function coefficient vanishes (after integrating out
the the fields ${\cal M}_{[\alpha\beta]}$ and $\Phi_\alpha$), but it is
still free in the infra-red.

\begin{center}
 {\em The $SU(2)\otimes Sp(6)\otimes U(1)_A$ Model}
\end{center}

{}This is the $\eta=1$ and $N=2$ model. This
model is interesting as supersymmetry is dynamically broken in this
model. The gauge group is $SU(2)\otimes Sp(6)\otimes U(1)_A$, the chiral matter
is given by $\Phi_\alpha=3\times ({\bf 3},{\bf 1})(+2)$ and 
$Q_\alpha=3\times ({\bf 2},
{\bf 6})(-1)$, and the tree-level superpotential is given by (\ref{tree}).  
The $U(1)_A$ D-term is given by
\begin{equation}
 D=2\Phi^2-Q^2+\xi_{\rm{\small FI}}~,
\end{equation}
where $\xi_{\rm{\small FI}}$ is negative for positive values of 
${\rm Im}(\phi_1)$.

{}First consider the case where $\Phi_\alpha=0$. Then the above D-term
vanishes only if $Q_\alpha=0$ and ${\rm Im}(\phi_1)=0$. The $SU(2)$ gauge
coupling is weak in the infra-red, so as far as the $Sp$ part of the gauge
theory is concerned we have $Sp(N_c=6)$ with $N_f=3$ flavors. In general,
the $Sp(N_c)$ theory with $N_f\leq N_c/2$ flavors (that is, $2N_f$ fields
$Q_i$, $i=1,\dots,2N_f$, in $N_c$ of $Sp(N_c)$) has a dynamically 
generated superpotential \cite{IP}:
\begin{equation}
 {\cal W}_{\rm{\small non-pert}}\sim 
 \left({\Lambda^{3(N_c/2+1)-N_f}\over {\rm Pf}(M)}\right)^{1/(N_c/2+1-N_f)}~,
\end{equation}
where $M_{ij}=-M_{ji}=Q_iQ_j$ are the meson fields, and $\Lambda$ is the
dynamically generated scale. For $N_f=N_c/2$ the gauge group is 
completely broken for ${\rm Pf}(M)\not=0$, and this superpotential is generated
by an instanton in the broken $Sp(N_c)$. For $N_f<N_c/2$ the
superpotential is associated with gaugino condensation in the unbroken
$Sp(N_c-2N_f)$. In particular, for $0<N_f\leq N_c/2$ the above superpotential
has a runaway behavior w.r.t. the vacuum expectation values of $M_{ij}$ (that
is, there is no supersymmetric vacuum for finite values of $M_{ij}$).  

{}In our case the mesons are ${\cal M}_{[\alpha\beta]}=3\times({\bf 3},
{\bf 1})(-2)$ and ${\cal M}_{\{\alpha\beta\}}=6\times({\bf 1},
{\bf 1})(-2)$. At first it might seem that we have global supersymmetry 
breaking as the $U(1)_A$ D-term grows with non-vanishing vacuum 
expectation values of the mesons. However, recall that we have assumed
that $\Phi_\alpha=0$. But there is nothing stopping $\Phi_\alpha$ from being 
non-vanishing. Suppose some or all $\Phi_\alpha\not=0$. Then the D-term
can {\em a priori} be set to zero. Nonetheless, we still have no
supersymmetric vacuum. Thus, in this case due to the tree-level superpotential
(\ref{tree}) two of the fields $Q_\alpha$ acquire masses (this is independent
of a particular configuration of the vacuum expectation values of $\Phi_\alpha$
as long as at least one of the fields $\Phi_\alpha\not=0$). So as far as the
$Sp(6)$ part of the gauge theory is concerned, at low energies we have $Sp(6)$
with one flavor. As we already mentioned above, in this theory we have a
dynamically generated runaway (in the corresponding meson field) superpotential
with no supersymmetric vacuum. The $D$-term, however, can now vanish, so
there is no stable vacuum with broken global supersymmetry.

{}Even so, as was discussed in detail in \cite{local}, generically we do expect
to have a stable vacuum with broken {\em local} supersymmetry. In particular,
if in the context of global supersymmetry we have a runaway superpotential
with the runaway directions corresponding to charged matter fields, then 
in the context of supergravity the runaway directions are generically expected
to be stabilized due to contributions coming from the K{\"a}hler potential
(see \cite{local} for details). Note that this mechanism is four-dimensional,
and we indeed have four-dimensional supergravity on D3-branes via the mechanism
of \cite{DG}.  

\subsection{Comments on the $\eta=-1$ Models}

{}In this subsection we would like to comment on some properties of the
${\bf Z}_3$ models with $\eta=-1$, that is, those where we have an 
O3$^-$-plane. As we have already mentioned, compact versions of some of these
models were discussed in \cite{Sagnotti,KaSh,KST,LPT,CVET}.

\begin{center}
 {\em The $SU(4)\otimes U(1)_A$ Model}
\end{center}

{}This is the $\eta=-1$ and $N=4$ model. The gauge group is $SU(4)\otimes
U(1)_A$, the chiral matter is given by $\Phi_\alpha=3\times {\bf 6}(+2)$, and
there is no tree-level superpotential. The $U(1)_A$ D-term is given by
\begin{equation}
 D=2\Phi^2+\xi_{\rm{\small FI}}~,
\end{equation}
where $\xi_{\rm{\small FI}}$ is negative for positive values of 
${\rm Im}(\phi_1)$.

{}The $SU(4)$ gauge theory is strongly coupled in the infra-red. We can view
this theory as $SO(6)$ with 3 vectors. In this theory there are two distinct
branches \cite{IS}. The first branch has a dynamically generated runaway 
superpotential in the vacuum expectation values of the mesons 
${\cal M}_{\{\alpha\beta\}}=6\times {\bf 1}(+2)$. On this branch supersymmetry
is broken via the mechanism mentioned at the end of the previous 
subsection.
The second branch has a vanishing superpotential, so supersymmetry is intact.
Note that in this model the orbifold blow-up mode can be non-zero. In fact, on
the first branch the non-supersymmetric vacuum has a non-zero blow-up mode,
while on the second branch it depends on the mesons 
${\cal M}_{\{\alpha\beta\}}$ (in both cases the blow-up mode is fixed from
the requirement that the $U(1)_A$ D-term vanish).

\begin{center}
 {\em The $SU(5)\otimes U(1)_A$ Model}
\end{center}

{}This is the $\eta=-1$ and $N=5$ model. The gauge group is $SU(5)\otimes
U(1)_A$, the chiral matter is given by $\Phi_\alpha=3\times {\bf 10}
(+2)$ and $Q_\alpha=3\times {\overline {\bf 5}}(-1)$, and
the tree-level superpotential is given by (\ref{tree}). The $U(1)_A$ D-term in
this model is given by
\begin{equation}
 D=2\Phi^2-Q^2+\xi_{\rm{\small FI}}~,
\end{equation}
where $\xi_{\rm{\small FI}}$ is negative for positive values of 
${\rm Im}(\phi_1)$.

{}To analyze the gauge dynamics in this model, let us first consider the
model with the gauge group $SU(5)$, chiral matter in $\Phi_\alpha=
3\times {\bf 15}$ and $Q_\alpha=3\times {\overline {\bf 5}}$, and {\em no}
tree-level superpotential. This theory is $s$-confining 
\cite{schmaltz}\footnote{The simplest example of an $s$-confining theory is
$SU(N_c)$ with $N_f=N_c+1$ flavors. In this theory we have confinement
without chiral symmetry breaking at the origin of the meson and baryon
moduli space.}. A simple way of understanding this is as follows. Consider
the following gauge invariant operator:
\begin{equation}
 \Sigma_{\{\alpha\beta\}}\equiv\Phi_{\{\alpha}\Phi_{\beta\}}
 \epsilon_{\gamma\delta\eta}\Phi_\gamma\Phi_\delta\Phi_\eta=6\times {\bf 1}~.
\end{equation}
So we have a D-flat direction corresponding to turning on non-vanishing
vacuum expectation values of $\Phi_\alpha$. The original $SU(5)$ gauge group
can be broken down to $SU(2)$ along this direction. To see this, consider
the branching of ${\bf 10}$ of $SU(5)$ under the breaking $SU(5)\supset
SU(3)\otimes SU(2)\otimes U(1)$:
\begin{eqnarray}
 &&{\bf 5}=({\bf 3},{\bf 1})(-2)+({\bf 1},{\bf 2})(+3)~,\\
 &&{\bf 10}=({\bf 1},{\bf 1})(+6)+({\overline {\bf 3}},{\bf 1})(-4)+
 ({\bf 3},{\bf 2})(+1)~. 
\end{eqnarray}
Now let us turn on non-zero vacuum expectation values for 
$({\bf 1},{\bf 1})(+6)$ in all
three $\Phi_\alpha$, and also for $({\overline {\bf 3}},{\bf 1})(-4)$ in
at least one of the $\Phi_\alpha$. This is consistent with the flat directions 
$\Sigma_{\{\alpha\beta\}}$. The gauge group is broken down to $SU(2)$, and
the left-over charged matter consists of three ${\bf 2}$'s coming from 
$\Phi_\alpha$ as well as three ${\bf 2}$'s coming from $Q_\alpha$. That is,
we have $SU(2)$ with 3 flavors, which is $s$-confining. 

{}Let us now include the tree-level superpotential (\ref{tree}). Then two of
the three ${\bf 2}$'s coming from $Q_\alpha$ acquire masses, and we have
$SU(2)$ with 2 flavors. In this theory we have quantum modification of the
moduli space \cite{Seiberg}, but supersymmetry is 
unbroken\footnote{In \cite{LPT} it was argued that supersymmetry is broken in
this model once we include the tree-level superpotential. In particular, the
non-perturbative superpotential
is known for the $SU(5)$ theory with $3\times{\bf 10}$ 
and $3\times{\overline{\bf 5}}$ and no tree-level superpotential 
\cite{schmaltz}. If we now
add the tree-level superpotential and write the total superpotential in
terms of the $SU(5)$ gauge invariant degrees of freedom, we will find that this
total superpotential has a runaway behavior in terms of the gauge invariant
degrees of freedom. This, however, does not necessarily 
imply that supersymmetry is 
broken. Thus, for vacuum expectation values of gauge invariant operators
larger than the dynamically generated scale $\Lambda$ in the original $SU(5)$
theory the description in terms of the $SU(5)$ gauge invariant operators
is no longer valid. Now, if supersymmetry were broken, the corresponding
non-supersymmetric vacuum would have various vacuum expectation values
stabilized at $\sim M_s$ (as this stabilization is due to the contributions 
coming from the K{\"a}hler potential as we discussed at the end of the previous
section), while in the case of a weakly coupled background ($g_s\ll 1$) we have
$\Lambda\ll M_s$. In this case we must therefore consider the low energy
theory after Higgsing (and not the other way around) as we did above where we
saw that supersymmetry is intact. However, if $g_s$ is somewhat large (the
one-loop $\beta$-function coefficient for the $SU(5)$ theory is 9), then
we could have supersymmetry breaking along the lines of \cite{LPT}.}. 
Finally,
note that the anomalous $U(1)_A$ D-term can also be canceled in this model.

\begin{center}
 {\em The $SU(6)\otimes SO(2)\otimes U(1)_A$ Model}
\end{center}

{}This is the $\eta=-1$ and $N=6$ model. The gauge group is $SU(6)\otimes
SO(2)\otimes
U(1)_A$, the chiral matter is given by $\Phi_\alpha=3\times ({\bf 15},{\bf 1})
(+2)$ and $Q_\alpha=3\times ({\overline {\bf 6}},{\bf 2})(-1)$, and
the tree-level superpotential is given by (\ref{tree}). (Note that $SO(2)\sim
U(1)$, and the doublet ${\bf 2}$ of $SO(2)$ refers to the states with 
opposite $U(1)$ charges.)  
The $U(1)_A$ D-term is given by
\begin{equation}
 D=2\Phi^2-Q^2+\xi_{\rm{\small FI}}~,
\end{equation}
where $\xi_{\rm{\small FI}}$ is negative for positive values of 
${\rm Im}(\phi_1)$. Note that 
\begin{equation}
 \Sigma\equiv\epsilon_{\alpha\beta\gamma}\Phi_\alpha\Phi_\beta\Phi_\gamma=
 ({\bf 1},{\bf 1})(+6)~.
\end{equation}
Thus, $\Sigma$ is a chiral gauge invariant operator w.r.t. $SU(6)\otimes
SO(2)$. So we have a D-flat direction, which is also F-flat, corresponding
to turning on non-vanishing expectation values of $\Phi_\alpha$ (the
$U(1)_A$ D-term can be canceled by appropriately turning on 
${\rm Im}(\phi_1)$). It is not difficult to see that 
at generic points along this flat direction the $SU(6)$ subgroup is broken
down to a $U(1)$, so the resulting gauge group is $U(1)\otimes SO(2)\sim
U(1)\otimes U(1)$. In the process of this Higgsing two of the original
three $Q_\alpha$ fields become massive, and we have total of 12 chiral
supermultiplets charged under this $U(1)\otimes U(1)$. These supermultiplets
can be used to Higgs the remaining Abelian gauge group completely.  

{}The above discussion suggests that there is no dynamically generated 
superpotential in this model. Another way of arriving at the same conclusion
is as follows. Consider giving vacuum expectation values to the fields 
$\Phi_\alpha$ so that $\Phi_1$ breaks $SU(6)$ down to $Sp(6)$, $\Phi_2$
breaks $Sp(6)$ down to $Sp(4)\otimes Sp(2)$, and finally $\Phi_3$ breaks
$Sp(4)\otimes Sp(2)$ down to $Sp(2)\otimes Sp(2)\otimes Sp(2)$ (this Higgsing 
is consistent with the flat direction $\Sigma$). The resulting gauge group is
$SU(2)\otimes SU(2)\otimes SU(2)\otimes SO(2)$ (the anomalous $U(1)_A$ 
is not shown), and the charged chiral matter
is given by 
$({\bf 2},{\bf 2},{\bf 1},{\bf 1})$,
$({\bf 2},{\bf 1},{\bf 2},{\bf 1})$,
$({\bf 1},{\bf 2},{\bf 2},{\bf 1})$,
$({\bf 2},{\bf 1},{\bf 1},{\bf 2})$,
$({\bf 1},{\bf 2},{\bf 1},{\bf 2})$,
$({\bf 1},{\bf 1},{\bf 2},{\bf 2})$.
We can further break the gauge group down to $SU(2)\otimes U(1)$ by giving 
non-vanishing expectation values to the last two fields. The resulting matter 
is given by $\chi^{\pm}_i={\bf 2}(\pm q_i)$, $i=1,2,3$, where $\pm q_i$ are
the $U(1)$ charges. As far as the $SU(2)$ subgroup is concerned, we have
$SU(2)$ with three flavors of quarks in the fundamental of $SU(2)$. In this
theory we have confinement without chiral symmetry breaking (at the origin of 
the moduli space), and there is no
non-perturbative superpotential \cite{Seiberg}. This suggests that the
$SU(6)\otimes SO(2)$ theory with the chiral matter $\Phi_\alpha$ and $Q_\alpha$
{\em and} the tree-level superpotential (\ref{tree}) is an 
$s$-confining theory. (Such theories {\em without} a tree-level superpotential
were classified in \cite{schmaltz}.)

\begin{center}
 {\em The $SU(N)\otimes SO(N-4)\otimes U(1)_A$ Models with $N\geq 7$}
\end{center}

{}These are the $\eta=-1$ and $N\geq 7$ models. The $SO(N-4)$ gauge
invariant operators are given by mesons
\begin{eqnarray}
 &&{\cal M}_{[\alpha\beta]}=3\times ({\overline {\bf A}},{\bf 1})(-2)~,\\
 &&{\cal M}_{\{\alpha\beta\}}=6\times ({\overline {\bf S}},{\bf 1})(-2)~, 
\end{eqnarray} 
and baryons
\begin{equation}
 \left({\cal B}_{\alpha_1\dots\alpha_{N-4}}\right)_{A_1\dots A_{N-4}}=
 Q_{\alpha_1 A_1 i_1}\cdots Q_{\alpha_{N-4} A_{N-4} i_{N-4}}
 \epsilon_{i_1\dots i_{N-4}}~,
\end{equation}
where $A_m$ is the ${\overline {\bf N}}$ index, while $i_m$ is the
${\bf N-4}$ index. (Note that the $U(1)_A$ charge of the baryon operators
is $-(N-4)$.)
There are no $SU(N)$ gauge invariant operators involving 
${\cal M}_{\{\alpha\beta\}}$. So all the $SU(N)$ gauge invariant operators
must be constructed from $\Phi_\alpha$, $\Theta_\alpha\equiv\epsilon_{\alpha
\beta\gamma}{\cal M}_{[\beta\gamma]}$ and the baryons
${\cal B}_{\alpha_1\dots\alpha_{N-4}}$. If $N$ is odd we have 
no gauge invariant operators involving totally antisymmetrized products of
only $\Phi_\alpha$ or only $\Theta_\alpha$. On the other hand, if $N$ is even
we have no such operators as $N/2>3$ (and the index $\alpha$ takes only three
values). So the building blocks for
the gauge invariant operators must be $\Phi_\alpha\Theta_\beta$ and
$\Phi_{\alpha_1}\cdots\Phi_{\alpha_n}{\cal B}_{\beta_1\dots\beta_{N-4}}$,
where $n=(N-4)/2$ if $N$ is even, and $n=N-2$ is $N$ is odd. This implies that
if $N$ is even all gauge
invariant operators have zero $U(1)_A$ charge. On the other hand, if $N$ is odd
the gauge invariant operators must have $U(1)_A$ charges which are positive
multiples of $N$.

{}Now, if 
$\Lambda$ is the dynamically generated scale of the $SU(N)$ theory, then
$\Lambda^{\beta_0}$ ($\beta_0=3N-{1\over 2}\times 3 \times (N-2)-
{1\over 2}\times 3\times (N-4)=9$ is the one-loop $\beta$-function coefficient
for $SU(N)$) has the $U(1)_A$ charge 
\begin{equation}
 q_A=(+2)\times{N(N-1)\over 2}+(-1)\times N\times (N-4)=3N~.
\end{equation} 
This together with the above arguments indicates 
that in theories with even $N$ we
cannot have a non-perturbative superpotential (as the superpotential must
have vanishing $U(1)_A$ charge). For odd $N$, however, this argument does not
rule out a non-perturbative superpotential.

{}To understand the odd $N$ cases in more detail, 
let us use the standard $U(1)_R$
symmetry arguments. We will assign $+1$ $U(1)_R$ charges to the gauginos
of the $SU(N)$ as well as $SO(N-4)$ gauge groups, and $U(1)_R$ charges 
$q_\Phi$ and $q_Q$ to the fields $\Phi_\alpha$ respectively $Q_\alpha$. Then
the requirement that the $U(1)_R SU(N)^2$ and $U(1)_R SO(N-4)^2$ anomalies
vanish gives the following values:
\begin{eqnarray}
 &&q_\Phi={{2N-12}\over 3N}~,\\
 &&q_Q={{2N+6}\over 3N}~.
\end{eqnarray} 
This implies that the $(U(1)_A,U(1)_R)$ charges read:
\begin{eqnarray}
 &&\Lambda^3:~~~(N,0)~,\\
 &&\Phi_\alpha\Theta_\beta:~~~(0,2)~,\\
 &&\Phi_{\alpha_1}\cdots\Phi_{\alpha_{N-2}}
 {\cal B}_{\beta_1\dots\beta_{N-4}}:~~~(N,2(2N-9)/3)~.
\end{eqnarray}
The superpotential must have $U(1)_R$ charge $+2$. We do have combinations
with $U(1)_R$ charge $+2$ (and $U(1)_A$ charge 0), which can schematically
be written as
\begin{equation}
 \left({\Phi_{\alpha_1}\cdots\Phi_{\alpha_{N-2}}
 {\cal B}_{\beta_1\dots\beta_{N-4}}\over \Lambda^3}\right)^{3\over{2N-9}}=
 \left({\left[\Phi_{\alpha_1}\cdots\Phi_{\alpha_{N-2}}
 {\cal B}_{\beta_1\dots\beta_{N-4}}\right]^3\over 
 \Lambda^9}\right)^{1\over{2N-9}}~.
\end{equation}
These combinations, however, are not holomorphic in the gauge invariant 
operators for odd $N\geq 7$ (in particular, we have a branch point at the
origin). We therefore conclude that for these values
of $N$ we do not have a non-perturbative superpotential either.

{}Next, recall that the tree-level superpotential is given by
\begin{equation}
 {\cal W}_{\rm{\small tree}}=y\Phi_\alpha\Theta_\alpha~. 
\end{equation}
The F-flatness conditions then imply that $\Theta_\alpha=0$ and
$\epsilon_{\alpha\beta\gamma}\Phi_{\beta[A_1A_2]} Q_{\gamma A_2 i}=0$.
In particular, in all cases at hand the gauge invariant operators involving
the $SO(N-4)$ mesons $\Theta_\alpha$ must vanish.
 
{}On the other hand, in the odd $N$ cases the
gauge invariant operators involving the $SO(N-4)$ baryons can be written as
\begin{equation}
 \left(\Phi_{\alpha_1}\Phi_{\alpha_2}\right)
 \left(\left(\Phi_{\beta_1}Q_{\gamma_1 i_1}\right)\cdots
 \left(\Phi_{\beta_{N-4}}Q_{\gamma_{N-4} i_{N-4}}\right)\right)
 \epsilon_{i_1\dots i_{N-4}}~.
\end{equation}
Due to the aforementioned F-flatness conditions the $\beta_m$ and $\gamma_m$
indices must be symmetrized pairwise. Each such symmetrization gives 6
of the $SO(3)$ global symmetry (note that $6=5+1$). So we have $(N-4)$ 
6's of $SO(3)$ completely antisymmetrized. However, at most 3 6's of
$SO(3)$ can be completely antisymmetrized without vanishing. This implies
that for $N>7$ the above gauge invariant operators should vanish to be
compatible with the F-flatness conditions, and the blow-up mode is
frozen at its vanishing value. So in the odd $N>7$ cases the $SO(N-4)$ gauge
subgroup is unbroken, and therefore so is the $SU(N)$ gauge subgroup.

{}A similar analysis can be performed in the even $N$ cases. Here the
gauge invariant operators involving the $SO(N-4)$ baryons can be written as
\begin{equation}
 \left(Q_{\alpha_1 i_1}\cdots Q_{\alpha_{(N-4)/2} i_{(N-4)/2}}\right)
 \left(\left(\Phi_{\beta_1}Q_{\gamma_1 j_1}\right)\cdots
 \left(\Phi_{\beta_{(N-4)/2}}Q_{\gamma_{(N-4)/2} j_{(N-4)/2}}\right)\right)
 \epsilon_{i_1\dots i_{(N-4)/2}j_1\dots j_{(N-4)/2}}~. 
\end{equation}
In this case we therefore have $(N-4)/2$ 6's of $SO(3)$ completely
antisymmetrized. This implies that for $N>10$ these gauge invariant
operators must vanish to be
compatible with the F-flatness conditions, and the blow-up mode is
frozen at its vanishing value. So in the even $N>10$ cases the $SO(N-4)$ gauge
subgroup is unbroken, and therefore so is the $SU(N)$ gauge subgroup.

{}Now, in the $N=7$ and $N=10$ cases we have 3 6's of $SO(3)$ completely
antisymmetrized. Since 6 is reducible ($6=5+1$), and a totally antisymmetric
product of 3 5's vanishes, we have $1\cdot 5 \cdot 5$ with the two 5's
antisymmetrized. In particular, the singlet $\Phi_\alpha Q_\alpha$ must be
non-vanishing, that is, for at least one value of $\alpha$ we must have
$\Phi_\alpha\not=0$ and $Q_\alpha\not=0$. Without loss of generality we can
choose $\Phi_1\not=0$ and $Q_1\not=0$. The F-flatness conditions then imply 
that either $\Phi_2=Q_2=0$ or $\Phi_3=Q_3=0$. But then the gauge invariant
operators containing $SO(N-4)$ baryons also vanish in the $N=7$ and $N=10$ 
cases.

{}Finally, consider the $N=8$ case, where we have 2 6's of $SO(3)$ 
antisymmetrized. In this case the relevant products are $1\cdot 5$ and
$5\cdot 5$ (the latter is antisymmetrized). So in this case the above
argument does not apply. However, for $N=8$ we have $SO(N-4)=SO(4)\sim
SU(2)_L\otimes SU(2)_R$, and the matter is
\begin{eqnarray}
 &&\Phi_\alpha=3\times ({\bf 28},{\bf 1},{\bf 1})(+2)~,\\
 &&L_\alpha=3\times ({\overline {\bf 8}},{\bf 2},{\bf 1})(-1)~,\\
 &&R_\alpha=3\times ({\overline {\bf 8}},{\bf 1},{\bf 2})(-1)~.
\end{eqnarray}
The basic $SU(2)_L\otimes SU(2)_R$ gauge invariant operators are
\begin{eqnarray}
 &&{\cal L}_\alpha=\epsilon_{\alpha\beta\gamma}L_\beta L_\gamma=3\times
 ({\overline {\bf 28}},{\bf 1},{\bf 1})(-2)~,\\
 &&{\cal L}_{\{\alpha\beta\}}=L_{\{\alpha} L_{\beta\}}=6\times
 ({\overline {\bf 36}},{\bf 1},{\bf 1})(-2)~,\\
 &&{\cal R}_\alpha=\epsilon_{\alpha\beta\gamma}R_\beta R_\gamma=3\times
 ({\overline {\bf 28}},{\bf 1},{\bf 1})(-2)~,\\
 &&{\cal R}_{\{\alpha\beta\}}=R_{\{\alpha} R_{\beta\}}=6\times
 ({\overline {\bf 36}},{\bf 1},{\bf 1})(-2)~
\end{eqnarray}
The mesons ${\cal L}_{\{\alpha\beta\}}$ and ${\cal R}_{\{\alpha\beta\}}$
cannot enter the $SU(8)$ gauge invariant operators, while the mesons
${\cal L}_\alpha={\cal R}_\alpha=0$ due to the F-flatness conditions.

{}The above arguments indicate that in the $N\geq 7$ models the gauge group
is unbroken, and the blow-up mode is zero 
(so that the $U(1)_A$ D-term vanishes).
At low energies the theory flows into an interacting fixed point. As was
pointed out in \cite{orient1}, in the large $N$ limit this superconformal
field theory is actually the same as the parent ${\cal N}=4$
supersymmetric $SO(3N-4)$ gauge theory (which, in turn, in the large
$N$ limit is the same as the parent ${\cal N}=4$ supersymmetric $SU(3N-4)$ 
gauge theory).

\section{Other Examples}

{}In this section we would like to briefly 
mention other interesting ${\bf Z}_p$
examples. In particular, we will discuss a ${\bf Z}_5$ example and a 
${\bf Z}_7$ example.

\subsection{A ${\bf Z}_5$ Example}

{}Consider the ${\bf Z}_5$ orbifold group whose generator $\theta$ has the
following action on the complex coordinates $z_\alpha$ ($\omega\equiv
\exp(2\pi i/5)$):
\begin{equation}
 \theta z_{1,2}=\omega z_{1,2}~,~~~\theta z_3=\omega^3 z_3~.
\end{equation}
The tadpole cancellation conditions have the following solution (see subsection
B of section III):
\begin{equation}
 {\rm Tr}\left(\gamma_{\theta^2}\right)=-4\eta 
 {1\over (1+\omega)^2(1+\omega^3)}=-4\eta(\omega+
 \omega^4)~.
\end{equation}
This then implies that
\begin{equation}
 \gamma_\theta={\rm diag}(I_N,\omega I_N,\omega^2 I_{N-4\eta},\omega^3
 I_{N-4\eta},\omega^4 I_N)~.
\end{equation}
In particular, consider the $\eta=-1$ and $N=0$ case. We then have
$SU(4)\otimes U(1)_A$ gauge group with a chiral supermultiplet in 
${\bf 6}(+2)$. As far as the non-Abelian part of the gauge group is
concerned, we can view this model as $SO(6)$ with one vector. In this theory
we have a runaway superpotential \cite{IS}, so supersymmetry is broken in
this model.

\subsection{A ${\bf Z}_7$ Example}

{}Consider the ${\bf Z}_7$ orbifold group whose generator $\theta$ 
has the following action on the complex coordinates $z_\alpha$ ($\omega\equiv
\exp(2\pi i/7)$):
\begin{equation}
 \theta z_1=\omega z_1~,~~~\theta z_2=\omega^2 z_2~~~\theta z_3=\omega^4 z_3~.
\end{equation}
The tadpole cancellation conditions have the following solution (see subsection
B of section III):
\begin{equation}
 \gamma_{\theta}=-4\eta~.
\end{equation}
This then implies that
\begin{equation}
 \gamma_\theta={\rm diag}(I_{N-4\eta},\omega I_N,\omega^2 I_N,\omega^3
 I_N,\omega^4 I_N,\omega^5 I_N,\omega^6 I_N)~.
\end{equation}
In particular, consider the $\eta=-1$ and $N=0$ case. We then have pure
$SO(4)\sim SU(2)_L\otimes SU(2)_R$ super-Yang-Mills theory. It would be 
interesting to use this setup to identify BPS domain walls in this gauge 
theory.

\section{Concluding Remarks}

{}The above discussions illustrate that the orientiworld framework has
a rich variety of non-perturbative phenomena that can arise in the
orientiworld models. In particular, we can have dynamical supersymmetry 
breaking as well as various interesting supersymmetry preserving phenomena
such as confinement, domain walls, {\em etc}. 

{}The orientiworld framework gives a consistent embedding of non-conformal
gauge theories in the Type IIB string theory context. Generalizing the
gauge/string theory correspondence of \cite{malda,GKP,witten} to such theories
would be very interesting, but it is also expected to be non-trivial as the
corresponding supergravity solutions are often expected to be singular.
Solving the problem of such singularities might also shed light on
non-supersymmetric cases, which would be interesting in the context of the
cosmological constant problem along the lines of \cite{DGS}.

\acknowledgments

{}This work was supported in part by the National Science Foundation and
an Alfred P. Sloan Fellowship. I would like to thank the New High
Energy Theory Center at Rutgers University for their kind hospitality while
parts of this work were completed.
I would also like to thank Albert and Ribena Yu for financial support.

\end{document}